\documentclass{emulateapj}
\newcommand{\beq}{\begin{equation}}
\newcommand{\eeq}{\end{equation}}
\newcommand{\etal}{{et~al. }}
\newcommand{\kms}{km~s$^{-1}$ }
\newcommand{\HPI}{7.43 $\pm$ 0.17 }
\accepted{}\shorttitle{Parallax of the Pleiades}\shortauthors{Soderblom et al.}
\begin{document}
\title{Confirmation of Errors in {\it Hipparcos} Parallaxes from HST/FGS Astrometry of the Pleiades\footnote{Based on observations made with the NASA/ESA Hubble Space Telescope, obtained at the Space Telescope Science Institute, which is operated by the Association of Universities for Research in Astronomy, Inc., under NASA contract NAS5-26555.}}
\author{David R. Soderblom and Ed Nelan}\affil{Space Telescope Science Institute\altaffilmark{2}, 3700 San Martin Drive, Baltimore, MD 21218}\email{soderblom, nelan@stsci.edu}
\author{G. Fritz Benedict, Barbara McArthur, Ivan Ramirez, and William Spiesman}\affil{McDonald Observatory, University of Texas, Austin, TX 78712 }\email{fritz, mca, ivan, spies@astro.as.utexas.edu}\and
\author{Burton F. Jones}\affil{Lick Observatory, University of California, Santa Cruz CA 95064}\email{jones@ucolick.org}\altaffiltext{2}{Operated by the Association of Universities for Research inAstronomy, Inc.\ under contract to the National Aeronautics and SpaceAdministration.}
\begin{abstract}
We present absolute trigonometric parallaxes and relative proper motions for three members of the Pleiades star cluster, obtained with {\it HST}'s Fine Guidance Sensor 1r, a white-light interferometer. We estimate spectral types and luminosity classes of the stars comprising the astrometric reference frame from $R \approx 2,000$ spectra, {\it VJHK} photometry, and reduced proper motions.  From these we derive estimates of absolute parallaxes and introduce them into our model as observations with error. We constrain the three cluster members to have a $1\sigma$ dispersion in distance less than 6.4 pc, and find an average $\pi_{abs}$ = \HPI $\pm$ 0.20 milliseconds of arc, where the second error is systematic due to member placement within the cluster. This parallax corresponds to a distance of $134.6\pm3.1$ pc or a distance modulus of $(m-M) = 5.65\pm0.05$ for these three Pleiads, presuming a central location. This result agrees with three other independent determinations of the Pleiades distance.  Presuming that the cluster depth systematic error can be significantly reduced because of the random placement of these many members within the cluster, these four independent measures yield a best-estimate Pleiades distance of $\pi_{abs} = 7.49\pm0.07$ milliseconds of arc, corresponding to a distance of $133.5\pm1.2$ pc or a distance modulus of $(m-M) = 5.63\pm0.02$. This resolves the dispute between the main sequence fitting and the {\it Hipparcos} distance moduli in favor of main sequence fitting.

\end{abstract}




%

\section{The Problem}

Our knowledge of the life histories of stars relies on models whose fidelity is ultimately tested by appeal to real stars. The Sun provides the most basic calibration of these models, of course, because it is only for the Sun that an accurate age exists and for which the mass, temperature, composition, and structure are known with precision, accuracy, and completeness.  Clusters of stars are also fundamental for constructing models because we can assume that all the cluster's members are of the same age and composition, even if other parameters are more loosely constrained. 

Preeminent among clusters is the Pleiades, and much effort has gone into determining the absolute parallax of this cluster. ESA's {\it Hipparcos} mission brought the benefits of space observing to astrometry to produce precise positions, proper motions, and parallaxes for nearly all stars brighter than $V \approx 9$.  Before {\it Hipparcos}, the distance to the Pleiades was too large for ground-based parallaxes to yield a good distance, so the best estimates were derived by comparing the main sequence of the Pleiades to a main sequence constucted from nearby stars with large parallaxes.  A small correction for evolution is necessary (the Pleiades is about 100 Myr old \citep{Pin98} while the nearby field stars are typically as old as the Sun), but the Pleiades appears to have essentially the same elemental abundances as the Sun \citep{Boe90}, obviating a need for a metallicity correction such as is needed, for example, for the Hyades.

In addition to its primary program, {\it Hipparcos} included stars in several of the nearest open clusters in order to resolve the ``Hyades distance problem" once and for all, and to similarly calibrate other clusters.  The result obtained by {\it Hipparcos} for the Pleiades \citep{vanL99} was a complete surprise, yielding a distance modulus of $(m-M) = 5.37\pm0.06$ magnitude, to be compared to a modulus of $5.60\pm0.04$ from main sequence fitting \citep{Pin98}.  Taken at face value, the {\it Hipparcos} result means that stars in the Pleiades are about 0.23 magnitude fainter than otherwise similar stars of the solar neighborhood.  This large discrepancy has forced a careful reexamination of the assumptions and input parameters of the stellar models, as well as a thorough study of the {\it Hipparcos} data itself and potential errors in it.  The controversy has not been fully resolved in that builders of star models find that the changes in physics or input parameters needed to account for the {\it Hipparcos} distance are too radical to be reasonable while the {\it Hipparcos} team has resolutely defended the {\it Hipparcos} result.  With no clear reconciliation of these divergent views, we felt it worthwhile to reobserve some stars in the Pleiades in the traditional method of parallax astrometry---highly precise measurements of stellar positions relative to nearby reference stars---by taking advantage of the extraordinary precision achieveable with Fine Guidance Sensor 1r on the {\it Hubble Space Telescope}.

This project began as an effort to resolve known Pleiades spectroscopic binaries into visual binaries so that we could both obtain an accurate distance and calibrate the Zero-Age Main Sequence with known masses.  We did not succeed in resolving the spectroscopic binaries, nor would our measurement of the Pleiades parallax by itself resolve the problem raised by {\it Hipparcos}, but our measurement in concert with other recent independent measurements of the Pleiades distance clearly and unambiguously shows that the {\it Hipparcos} parallax is wrong and that traditional main sequence fitting results in reliable estimates.  To avoid repetition, we will discuss the work to date in detail in our discussion.

\section{Observations and Data Reduction}  \label{AstRefs}
\begin{deluxetable}{clcccc}
\tablewidth{0in}
\tablecaption{Pleiades Log of Observations and Member Photometry\label{tbl-LOO}}
\tablehead{
\colhead{Set}&\colhead{MJD}&\colhead{Roll \tablenotemark{a}}&\colhead{}&\colhead{$V$\tablenotemark{b}} &\colhead{} \\
\colhead{}&\colhead{}&\colhead{(degrees)}&\colhead{3030}&\colhead{3063}&\colhead{3179}
}
\startdata
1 & 51770.65507 &  284.046 & 14.03 & 13.54 & 10.05  \\
2 & 51783.6811   &  284.046 & 14.00 & 13.56 & 10.06  \\
3 & 51957.17546 &  103.014 & 14.00 & 13.47 & 10.08  \\
4 & 51968.37565 &  103.014 & 14.01 & 13.57 & 10.09  \\
5 & 52128.77383 &  284.046 & 13.97 & 13.58 & 10.07  \\
6 & 53053.24519 &  113.021 & 14.00 & 13.44 & 10.08  \\
 ~\\
$\langle V\rangle$ & & & 14.00 & 13.53 & 10.07  \\
$\sigma_V$ & & & 0.02 & 0.06 & 0.01  \\
\enddata
\tablenotetext{a}{Spacecraft roll as defined in Chapter 2, FGS Instrument Handbook \citep{Nel01}}
\tablenotetext{b}{Average of 2 to 5 observations at each epoch. Internal errors are on order 0.005 magnitude per observation set.
}
\end{deluxetable}

\begin{figure}
\plotone{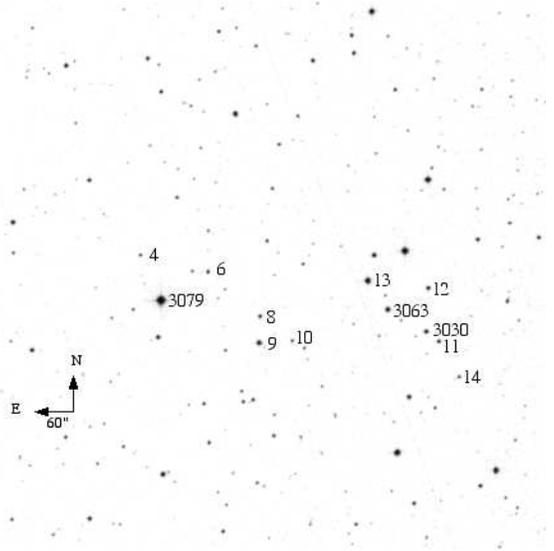}
\caption{The Pleiades cluster, its members, and the astrometric reference stars observed. }
\label{fig-1}
\end{figure}
Six sets of astrometric data were acquired with {\it HST}, spanning 3.51 years, for a total of 135 measurements  of the three Pleiads and nine reference stars.  The three Pleiades targets were H {\sc ii} 3030, 3063, and 3179, identified hereafter by their \citet{Her47} numbers. Table~\ref{tbl-LOO} lists the epochs of observation and measured FGS $V$-band photometry of the three Pleiads. Each data set required approximately 33 minutes of spacecraft time. The reductions and calibrations are detailed in \cite{Ben02a},  \cite{Ben02b}, and \cite{mca01}.  At each epoch we measured both the reference stars and the target multiple times in order to to correct for intra-orbit drift of the type seen in the cross filter calibration data shown in Figure 1 of \cite{Ben02a}. Figure \ref{fig-1} shows the distribution of the reference stars (4--14) and the presumed Pleiads (3030, 3063, and 3179) on a second-generation $R$-band image, obtained from the Digital Sky Survey (http://stdatu.stsci.edu/dss/).

\citet{Bra91} and \citet{Nel01} provide an overview of {\it HST}'s Fine Guidance Sensors, and \citet{Ben02b} describe the fringe tracking (POS) mode astrometric capabilities of an FGS, along with data acquisition and reduction strategies also used in the present study.  Times of observation use a modified Julian Date, MJD  =  JD $-$ 2444000.5.

We obtained observations at each of the two maximum parallax factors.  This leads to the two distinct spacecraft roll angles shown which result from the requirement to keep {\it HST}'s solar panels fully illuminated throughout the year.  This roll constraint generally imposes alternate orientations at each time of maximum positive or negative parallax factor over a typical 2.5 year parallax campaign, allowing a clean separation of parallax and proper motion signatures. As noted, our original intent was to determine orbital parameters for some known spectroscopic binaries, but once resolution of the binary did not work out we changed this dynamical parallax experiment to a standard parallax program. The most recent data set extended our time span by 2.5 years, significantly improving the accuracy of our final parallaxes and the precision of our final proper motion values.
\setcounter{footnote}{0}

\section{Absolute Parallaxes for the Reference Stars} \label{SpecPhot}

Because the parallax determined for the three Pleiades members is measured with respect to reference frame stars which have their own parallaxes, we must either apply a statistically-derived correction from relative to absolute parallax (\citet{Alt95}, hereafter YPC95), or estimate the absolute parallaxes of the reference frame stars.  In principle, the colors, spectral type, and luminosity class of a star can be used to estimate the absolute magnitude, $M_V$, and $V$-band absorption, $A_V$. The absolute parallax is then simply,
\beq
\pi_{\rm abs} = 10^{-(V-M_V+5-A_V)/5}
\eeq

The luminosity class is generally more difficult to estimate than the spectral type (temperature class), yet the derived absolute magnitudes are critically dependent on the assumed luminosity. As a consequence, we use as much additional information as possible in an attempt to confirm the luminosity classes. Specifically, we obtained 2MASS photometry and UCAC2 proper motions for a one-degree-square field containing Figure \ref{fig-1}, and then iteratively employ the technique of reduced proper motion (\citet{Yon03}, \citet{Gou03}) in an effort to discriminate between giants and dwarfs.

\subsection{ Reference Star Photometry}
\begin{deluxetable*}{cccccc}
\tablewidth{0in}
\tablecaption{FGS and Near-IR Photometry \label{tbl-IR}}
\tablehead{\colhead{ID}&
\colhead{$V$} &
\colhead{$K$} &
\colhead{$(J-H)$} &
\colhead{$(J-K)$} &
\colhead{$(V-K)$}
}
\startdata
3179  & 10.07 &   $8.68\pm0.02$ & $0.32\pm0.02$ & $0.37\pm0.02$ & $1.40\pm0.10$  \\
3063  & 13.54 & $10.34\pm0.02$ & $0.67\pm0.03$ & $0.84\pm0.03$ & $3.20\pm0.10$  \\
3030  & 14.00 & $10.63\pm0.02$ & $0.71\pm0.03$ & $0.85\pm0.03$ & $3.37\pm0.10$  \\
ref-4   & 15.66 & $13.98\pm0.05$ & $0.43\pm0.05$ & $0.49\pm0.06$ & $1.70\pm0.11$  \\
ref-6   & 14.56 & $12.03\pm0.02$ & $0.56\pm0.02$ & $0.68\pm0.03$ & $2.54\pm0.10$  \\
ref-8   & 14.48 & $12.91\pm0.03$ & $0.29\pm0.03$ & $0.37\pm0.04$ & $1.57\pm0.10$  \\
ref-9   & 13.60 & $10.64\pm0.02$ & $0.68\pm0.02$ & $0.79\pm0.03$ & $2.97\pm0.10$  \\
ref-10 & 15.85 & $13.40\pm0.04$ & $0.55\pm0.04$ & $0.67\pm0.05$ & $2.45\pm0.11$  \\
ref-11 & 14.63 & $12.75\pm0.02$ & $0.43\pm0.03$ & $0.49\pm0.04$ & $1.88\pm0.10$  \\
ref-12 & 14.23 & $12.15\pm0.03$ & $0.48\pm0.04$ & $0.51\pm0.04$ & $2.10\pm0.10$  \\
ref-13 & 13.15 & $10.57\pm0.02$ & $0.34\pm0.03$ & $0.42\pm0.03$ & $1.57\pm0.10$  \\
ref-14 & 15.48 & $13.78\pm0.03$ & $0.42\pm0.05$ & $0.44\pm0.05$ & $1.70\pm0.11$  \\
\enddata
\end{deluxetable*}

Our bandpasses for reference star photometry include:  $V$ (from FGS 1r), and {\it JHK} (from 2MASS\footnote{The Two Micron All Sky Survey is a joint project of the University of Massachusetts and the Infrared Processing and Analysis Center/California Institute of Technology }).  The 2MASS {\it JHK} values have been transformed to the \citet{Bes88} system using the transformations provided in \citet{Car01}. Table \ref{tbl-IR} lists {\it VJHK} photometry for the target and reference stars indicated in Figure~\ref{fig-1}.

\subsection{Reference Star Spectroscopy }

The spectra from which we estimated spectral type and luminosity class come from Lick Observatory\footnote{Lick Observatory is owned and operated by the University of California.}. The resolution was approximately, with coverage from 3900 \AA~ to 6700 \AA. Classifications used a combination of template matching and line ratios. Spectral types for the stars are good to about 2
subclasses. Table \ref{tbl-SPP} lists the spectral types and luminosity classes for our reference stars. The estimated classification uncertainties are used to generate the $\sigma_{M_V}$ values in that table. 

\subsection{Interstellar Extinction} \label{AV}
\begin{figure}
\plotone{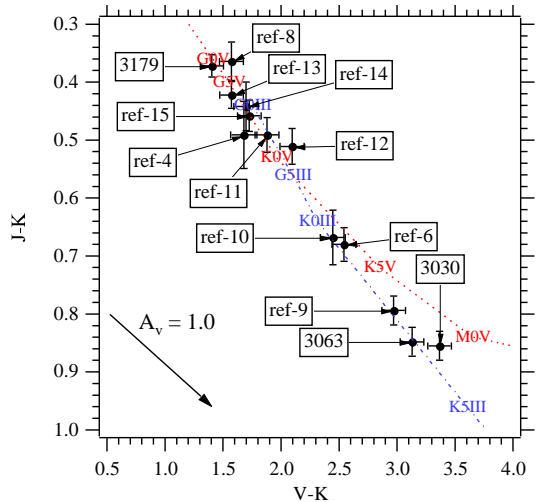}
\caption{$(J-K)$ vs. $(V-K)$ color-color diagram. The dashed line is the locus of  dwarf (luminosity class V) stars of various spectral types; the dot-dashed line is for giants (luminosity class III). The reddening vector indicates $A_V = 1.0$ for the plotted color systems.}
\label{fig-2}
\end{figure}
To determine interstellar extinction, we first plot these stars in a $(J-K)$ vs. $(V-K)$ diagram. A comparison of the relationships between spectral type and intrinsic color against those we measured provides an estimate of reddening. Figure \ref{fig-2} shows this color-color diagram and a reddening vector for $A_V = 1.0$. Also plotted are mappings between spectral type and luminosity classes V and III from \citet{Bes88} and \citet{Cox00} (hereafter AQ2000).  Figure~\ref{fig-2}, along with the estimated spectral types, provides an indication of the reddening for each reference star. 

\begin{deluxetable}{llcccc}
\tablewidth{0in}
\tablecaption{Reference Star $A_V$ from Spectrophotometry  \label{tbl-AV}}
\tablehead{  \colhead{ID}&
\colhead{SpT}&   \colhead{$(V-K)_0$}&  \colhead{$(V-K)$} &  \colhead{$E(V-K)$} &
\colhead{$A_V$\tablenotemark{a}}
}
\startdata
ref-4   & G5V  & 1.55 & 1.68 & 0.13 & 0.14 \\
ref-6   & K1IV  & 2.32 & 2.54 & 0.22 & 0.24 \\
ref-8   & G3V  & 1.45 & 1.57 & 0.12 & 0.13 \\
ref-9   & K2III   & 2.70 & 2.97 & 0.27 & 0.30 \\
ref-10 & K1IV  & 2.32 & 2.45 & 0.13 & 0.14 \\
ref-11 & G8V  & 1.80 & 1.88 & 0.08 & 0.09 \\
ref-12 & K0V   & 1.96 & 2.10 & 0.13 & 0.15 \\
ref-13 & G3V   & 1.45 & 1.57 & 0.12 & 0.13 \\
ref-14 & G5V   & 1.55 & 1.70 & 0.15 & 0.16 \\
\enddata
\tablenotetext{a}{$A_V = 1.1E(V-K)$}
\end{deluxetable}
\begin{deluxetable}{cclccrr}
\tablewidth{0in}
\tablecaption{Astrometric Reference Star Spectral Classifications and
Spectrophotometric Parallaxes \label{tbl-SPP}}
\tablehead{\colhead{ID}& \colhead{Sp. T.}&
\colhead{V} & \colhead{M$_V$} & \colhead{A$_V$} &\colhead{m-M}&
\colhead{$\pi_{abs}$(mas)}}
\startdata
ref-4   & G5V  & 15.68 & 5.1 & 0.14 & $10.6\pm0.7$ & $0.8\phn\pm0.3$  \\
ref-6   & K1IV & 14.5    & 3.4 & 0.23 & $11.2\pm2\hphantom{.7}$    & $0.06\pm0.6$  \\
ref-8   & G3V  & 14.48 & 4.8 & 0.14 & $9.7\pm0.7$   & $1.2\phn\pm0.4$  \\
ref-9   & K2III  & 13.61 & 0.5 & 0.23 & $13.1\pm0.7$ & $0.3\phn\pm0.1$  \\
ref-10 & K1IV & 15.85 & 3.4 & 0.23 & $12.5\pm2\hphantom{.7}$    & $0.4\phn\pm0.3$  \\
ref-11 & G8V & 14.63  & 5.6 & 0.14 & $9.1\pm0.7$   & $1.7\phn\pm0.5$  \\
ref-12 & K0V & 14.24  & 5.9 & 0.14 & $8.3\pm0.7$   & $2.3\phn\pm0.7$  \\
ref-13 & G3V & 12.14  & 4.8 & 0.14 & $7.3\pm0.7$   & $3.66\pm1.2$  \\
ref-14 & G5V & 15.48  & 5.1 & 0.14 & $10.4\pm0.7$ & $0.9\phn\pm0.3$  \\
\enddata
\end{deluxetable}

Assuming an $R = 3.1$ Galactic reddening law \citep{Sav79}, we derive $A_V$ values by comparing the measured colors (Table \ref{tbl-IR}) with intrinsic $(V-K)_0$ colors from \citet{Bes88} and AQ2000.  Specifically, we estimate $A_V$ from $A_V/E(V-K) = 1.1$, derived from the \citet{Sav79} reddening law. The resulting $A_V$ values are collected in Table \ref{tbl-AV}. Colors and spectral types of these reference stars are consistent with a field-wide average $\langle A_V\rangle =0.17\pm0.06$, far less than the maximum reddening, $A_V < 0.72$, determined by \citet{Sch98}. For the stars classified as dwarfs, $\langle A_V \rangle =0.14\pm0.03$, in good agreement with a recent determination of $A_V$ = 0.12 for the Pleiades \citep{Hai01}.  The more distant non-dwarfs have $\langle A_V \rangle =0.23\pm0.08$.

The technique of reduced proper motions can confirm the reference stars' estimated luminosity classes, but the precision of existing proper motions for all the reference stars was so low that only suggestive discrimination between giants and dwarfs was possible. Typical uncertainties for $H_K$, a parameter equivalent to absolute magnitude, $M_V$, were about a magnitude. Nonetheless, a reduced proper motion diagram does suggest that ref-6, -9, and -10 are not dwarf stars, since they are considerably redder in $(J-K)$ than the other stars classified as dwarfs. Giants are typically redder in $(J-K)$ than dwarfs for a given spectral type (AQ2000). Our luminosity class uncertainty is reflected in their input spectrophotometric parallax errors (Table~\ref{tbl-SPP}). We will revisit this additional test in Section~\ref{ASTMOD}, once we have solved for higher precision proper motions. 

\subsection{Adopted Reference Frame Absolute Parallaxes}

We derive absolute parallaxes with $M_V$ values from AQ2000 and the $\langle A_V \rangle$ derived from the photometry. Our adopted errors for $(m-M)_0$ are 0.7 mag for the dwarfs and from 0.7 to 2 mag for the non-dwarf reference stars. These are somewhat larger than we have used in the past (\citet{Ben02a}, \citet{Ben02b}, \citet{McA02}), but justified given our far smaller set of spectrophotometric data. Our parallax values are listed in Table \ref{tbl-SPP}. Individually, no reference star absolute parallax is better determined than ${\sigma_{\pi}\over \pi}$ = 32\%. The average absolute parallax for the reference frame is $\langle\pi_{abs}\rangle = 1.3$ mas.  As a check, we compare this to the correction to absolute parallax discussed and presented in YPC95 (Sec.~3.2, Fig.~2).  Entering YPC95, Fig.~2, with the Pleiades Galactic latitude, $b = -23\fdg0$, and average magnitude for the reference frame, $\langle V_{\rm ref} \rangle = 14.5$, we obtain a correction to absolute of 1.0 mas. We prefer to introduce into our reduction model our spectrophotmetrically-estimated reference star parallaxes as observations with error. When such data are available, the use of spectrophotometric parallaxes offers a more direct (i.e., less Galaxy model-dependent) way of determining the reference star absolute parallaxes.

\section{The Absolute Parallax of the Pleiades}

\subsection{The Astrometric Model} \label{ASTMOD}

Using the positions measured by FGS 1r, we determine the scale, rotation, and offset ``plate constants" relative to an arbitrarily-adopted constraint epoch (the so-called ``master plate") for each observation set (the data acquired at each epoch). The MJD of each observation set is listed in Table~\ref{tbl-LOO}, along with a measured magnitude transformed from the FGS instrumental system as per \citet{Ben98}.  Our Pleiades reference frame contains 9 stars.  We employ the six-parameter model discussed in \citet{Ben99} for those observations. In this case, we determined the plate parameters from target and reference star data. Additionally, we apply corrections for lateral color discussed in \citet{Ben99}, using values specific to FGS 1r as determined from observations with that FGS.

As for all our previous astrometric analyses, we employ GaussFit (\citet{Jef87}) to minimize $\chi^2$. The solved equations of condition for the Pleiades field are:
\beq
        x^\prime = x + lcx(\it B-V)
\eeq
\beq
        y^\prime = y + lcy(\it B-V)
\eeq
\beq
\xi = Ax^\prime + By' + C  - \mu_x \Delta t  - P_\alpha\pi_x
\eeq
\beq
\eta = -Bx^\prime + Ay^\prime + F  - \mu_y \Delta t  - P_\delta\pi_y
\eeq

\noindent
where $\it x$ and $\it y$ are the measured coordinates from {\it HST}; $\it lcx$ and $\it lcy$ are the lateral color corrections; and $(B-V)$ represents the $(B-V)$ color of each star, estimated from its spectral type, $A_V$, and $(J-K)$ color listed in Table~\ref{tbl-IR}.  $A$ and $B$ are scale- and rotation plate constants, $C$ and $F$ are offsets; $\mu_x$ and $\mu_y$ are proper motions; $\Delta t$ is the epoch difference from the mean epoch; $P_\alpha$ and $P_\delta$ are parallax factors;  and $\it \pi_x$ and $\it \pi_y$ are  the parallaxes in $x$ and $y$.  We obtain the parallax factors from a JPL Earth orbit predictor \citep{Sta90}, upgraded to version DE405. Orientation to the sky is obtained from ground-based astrometry (2MASS Catalog) with uncertainties in the field orientation of $0\fdg05$.

\subsection{Modeling Constraints from Prior Knowledge} \label{Priors}

In addition to introducing our estimated reference star parallaxes as observations with error, we also introduce proper motion data from UCAC2 (\citet{Zac03}) and \citet{Sch95}. Initial values are listed in Table~\ref{tbl-PM}. Typical input errors are 5-6 mas for each coordinate. The lateral color calibrations and the $(B-V)$ color indices are also treated as observations with error. As a final constraining observation, we solve for a line-of-sight dispersion in the parallaxes of the three Pleiades members with the `observation' derived from the $1\sigma$ angular extent of the Pleiades (1\arcdeg, from Adams \etal 2001) and an assumption of spherical symmetry. From this, we infer that the 1$\sigma$ dispersion in distance in this group is 1\arcdeg/1 radian =1.7\%.  Hence, the 1$\sigma$ dispersion in the parallax difference between Pleiades
members  is
\beq
\Delta \pi  =  1.7\% \times \sqrt{2} \times 7.7~ {\rm mas} = 0.20 ~{\rm mas,}
\eeq
where we have here temporarily adopted a parallax of the Pleiades, $\langle \pi \rangle =  7.7$ mas.  The parallax dispersion among targets 3030, 3179, and 3063 becomes an observation with associated error fed to our model, an observation used to estimate the parallax dispersion among the three stars, while solving for their parallaxes. Loosening the cluster $1\sigma$ dispersion to 2\arcdeg\ (i.e., $\Delta \pi  = 0.38$ mas) had no effect on the final weighted average parallax. Again, note that $\Delta \pi  = 0.2$ mas is not an error associated with the {\em distance} to the Pleiades. It serves to constrain the dispersion in distances measured for Pleiades members.

\begin{figure}
\plotone{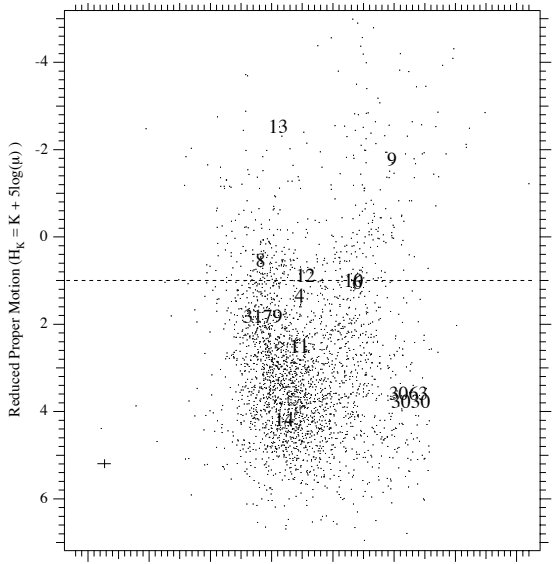}
\caption{Reduced proper motion diagram for 5,542 stars in a one-degree field centered on the Pleiades. Star identifications are shown for our Pleiades targets (H {\sc ii} 3030,3063, and 3179) and our astrometric reference stars. For a given spectral type, giants and sub-giants have more negative $H_K$ values and are redder than dwarfs in $(J-K)$.  Stars ref-6 and ref-10 are coincident.  $H_K$ values are derived from `Final' proper motionsin Table \ref{tbl-PM}. The small cross at the lower left represents a typical $(J-K)$ error of 0.04 mag and $H_K$ error of 0.17 mag. The horizontal dashed line is a giant-dwarf demarkation derived from a statistical analysis of the Tycho input catalog(Ciardi 2004, private communication).} \label{fig-3}
\end{figure}

Proper motion values obtained from our modeling of {\it HST} data are listed in Table~\ref{tbl-PM} as `Final'. We now employ the technique of reduced proper motions to provide a confirmation of the reference star estimated luminosity class listed in Table~\ref{tbl-SPP}. We obtain proper motion and $J$, $K$ photometry from UCAC2 and 2MASS for a one-degree-square field centered on the Pleiades. Figure~\ref{fig-3} shows $H_K = K + 5\log(\mu)$ versus $(J-K)$ color index for 5,542 stars. If all stars had the same transverse velocities, Figure~\ref{fig-3} would be equivalent to an H-R diagram. Target Pleiads and reference stars are plotted as ID numbers from Table~\ref{tbl-PM}. Errors in $H_K$ are now $\sim0.3$ mag. Reference stars 6, 9, and 10 are clearly separated from the others, supporting their classification as non-dwarfs. Ref-6 and ref-10 remain below ref-9, confirming their sub-giant nature.

\subsection{Assessing Reference Frame Residuals}

\begin{figure}
\epsscale{0.75}
\plotone{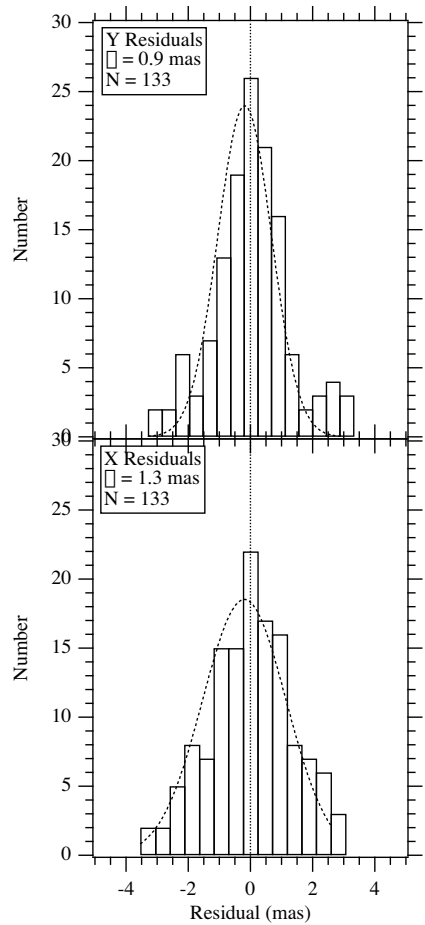}
\caption{Histograms of $x$ and $y$ residuals obtained from modeling the Pleiades members and astrometric reference stars with equations 4 and 5. Distributions are fit with gaussians whose $1\sigma$ dispersions are noted in the plots.} \label{fig-4}
\end{figure}

\begin{deluxetable}{ccrr}
\tablewidth{0in}
\tablecaption{Pleiades and Reference Star Positions    \label{tbl-POS}}
\tablehead{\colhead{ID}&
\colhead{$V$} &
\colhead{$\xi$ \tablenotemark{a}} &
\colhead{$\eta$ \tablenotemark{a}}
}
\startdata
3179  & $10.08$ & $+163.1991\pm0.0002$  & $-13.1667\pm0.0002$  \\
3063  & $13.47$ & $-198.8822\pm0.0003$ & $+59.7779\pm0.0003$  \\
3030  & $14.00$ & $-268.2952\pm0.0003$ & $+39.2897\pm0.0003$  \\
ref-4   & $15.68$ & $+213.0875\pm0.0012$  & $+51.9487\pm0.0010$  \\
ref-6   & $14.57$ & $+99.5992\pm0.0003$    & $+51.2020\pm0.0006$  \\
ref-8\tablenotemark{b} & $14.47$ & $0.0000\pm0.0005$ & $0.0000\pm0.0006$  \\
ref-9   & $13.61$ & $-8.4395\pm0.0003$     & $-42.3189\pm0.0003$  \\
ref-10 & $15.85$ & $-60.3818\pm0.0009$   & $-26.3444\pm0.0008$  \\
ref-11 & $14.63$ & $-292.6740\pm0.0003$ & $+28.8663\pm0.0004$  \\
ref-12 & $14.24$ & $-254.6792\pm0.0004$ & $+109.6633\pm0.0005$  \\
ref-13 & $12.14$ & $-156.6322\pm0.0003$ & $+97.8026\pm0.0003$  \\
ref-14 & $15.48$ & $-338.2213\pm0.0006$ & $-19.8165\pm0.0007$  \\
\enddata
\tablenotetext{a}{$\xi$ and $\eta$ are relative positions in arcseconds}
\tablenotetext{b}{RA = 03 51 45.050, Dec = +23 53 43.43, J2000}
\end{deluxetable}

Our initial modeling attempts indicated that three of the original twelve reference stars exhibited significantly larger residuals than average. These reference stars  were near the top- and bottom edges of the FGS1r field of regard. The Optical Field Angle Distortion calibration (\citet{McA02}) reduces as-built {\it HST} telescope and FGS1r distortions with amplitude $\sim1\arcsec$ to below 2 mas over much of the FGS1r field of regard. However, because the fidelity of correction drops precipitously near the edge of the field of regard, we removed these three stars from the solution. From histograms of the remaining reference star astrometric residuals (Figure~\ref{fig-4}) we conclude that we have obtained satisfactory correction. The resulting reference frame `catalog' in $\xi$ and $\eta$ standard coordinates (Table \ref{tbl-POS}) was determined with $\langle \sigma_\xi \rangle = 0.5$ and $\langle \sigma_\eta \rangle = 0.5$ mas.

To determine if there might be unmodeled---but possibly correctable---systematic effects at the 1 mas level, we plotted the Pleiades reference frame $x$ and $y$ residuals against a number of spacecraft, instrumental, and astronomical parameters. These included $x$ and $y$ position within the FGS ``pickle"; radial distance from the pickle center; reference star $V$ magnitude and $(B-V)$ color; and epoch of observation.  We saw no obvious trends, other than an expected increase in positional uncertainty with reference star magnitude.

\subsection{The Absolute Parallax of The Pleiades} \label{AbsPi}
\begin{deluxetable*}{ccrccc}
\tablewidth{0in}
\tablecaption{Pleiades and Reference Star Proper Motions \label{tbl-PM}}
\tablehead{\colhead{}&
\colhead{}&
\colhead{Input (UCAC2)} &
\colhead{}&
\colhead{Final ({\it HST})} &
\colhead{}  \\
\colhead{ID}&
\colhead{V} &
\colhead{$\mu_x$\tablenotemark{a}} &
\colhead{$\mu_y$\tablenotemark{a}}&
\colhead{$\mu_x$\tablenotemark{a}} &
\colhead{$\mu_y$\tablenotemark{a}} }
\startdata 
3179  & 10.08 & $+0.0192\pm0.0006$ & $-0.0465\pm0.0006$ & $+0.0192\pm0.0003$ & $-0.0465\pm0.0002$  \\
3063\tablenotemark{b} & 13.47 & $+0.0164\pm0.0011$ & $-0.0418\pm0.0011$ & $+0.0168\pm0.0003$ & $-0.0421\pm0.0005$  \\
3030\tablenotemark{b} & 14.00 & $+0.0154\pm0.0004$ & $-0.0408\pm0.0004$ & $+0.0155\pm0.0004$ & $-0.0403\pm0.0002$  \\
ref-4 & 15.68 & $+0.0035\pm0.0056$ & $-0.0034\pm0.0056$ & $+0.0024\pm0.0026$ & $-0.0019\pm0.0024$  \\
ref-6 & 14.57 & $+0.0054\pm0.0056$ & $-0.0094\pm0.0056$ & $+0.0044\pm0.0010$ & $-0.0049\pm0.0010$  \\
ref-8 & 14.47 & $+0.0052\pm0.0056$ & $-0.0119\pm0.0056$ & $+0.0030\pm0.0005$ & $-0.0016\pm0.0005$  \\
ref-9 & 13.61 & $+0.0117\pm0.0056$ & $+0.0033\pm0.0056$ & $-0.0014\pm0.0015$ & $+0.0031\pm0.0019$  \\
ref-10 & 15.85 & $-0.0016\pm0.0069$ & $-0.0092\pm0.0069$ & $-0.0016\pm0.0022$ & $-0.0091\pm0.0026$  \\
ref-11 & 14.63 & $+0.0030\pm0.0056$ & $-0.0131\pm0.0056$ & $-0.0043\pm0.0007$ & $-0.0039\pm0.0006$  \\
ref-12 & 14.24 & $+0.0058\pm0.0056$ & $-0.0096\pm0.0056$ & $-0.0008\pm0.0010$ & $+0.0024\pm0.0012$  \\
ref-13 & 12.14 & $-0.0074\pm0.0019$ & $-0.0129\pm0.0019$ & $-0.0093\pm0.0010$ & $-0.0083\pm0.0011$  \\
ref-14 & 15.48 & $0.0000\pm0.0058$ & $-0.0053\pm0.0058$ & $-0.0056\pm0.0024$ & $-0.0051\pm0.0032$ \\ 
\enddata
\tablenotetext{a}{$\mu_x$ and $\mu_y$ are relative motions in arcsec
yr$^{-1}$ }\tablenotetext{b}{$\mu_x$ and $\mu_y$ from \cite{Sch95}. }
\end{deluxetable*}

\begin{deluxetable}{crcc}
\tablewidth{0in}
\tablecaption{Pleiades and Reference Star Parallaxes and Transverse Velocities\label{tbl-PiVt}}
\tablehead{
\colhead{ID}&
\colhead{$\mu$\tablenotemark{a}} &
\colhead{$\pi_{abs}$\tablenotemark{b}} &
\colhead{$V_t$\tablenotemark{c}}
 \\
\colhead{}&
\colhead{mas yr$^{-1}$}&
\colhead{mas}&
\colhead{\kms}
}
\startdata
3179 & $50.36\pm0.40$ & $7.45\pm0.16$ &32\\
3063 & $45.30\pm0.53$ & $7.43\pm0.16$ &29\\
3030 & $43.20\pm0.48$ & $7.41\pm0.18$ &28\\
4        & $3.07\pm3.54$   & $0.82\pm0.09$ &18\\
6        & $6.63\pm1.39$   & $0.84\pm0.25$ &38\\
8        & $3.40\pm0.76$   & $1.21\pm0.13$ &13\\
9        & $3.42\pm2.43$   & $0.26\pm0.03$ &61\\
10     & $9.28\pm3.39$    & $0.36\pm0.11$ &\llap{1}22\\
11     & $5.79\pm0.98$    & $1.66\pm0.16$ &17\\
12     & $2.50\pm1.52$    & $2.25\pm0.23$ &\phn5\\
13     & $12.50\pm1.46$  & $1.64\pm0.32$ &36\\
14     & $7.58\pm4.04$    & $0.92\pm0.10$ &39\\
\enddata
\tablenotetext{a}{ $\mu = (\mu_x^2 +\mu_y^2)^{1/2}$ from `Final' in Table \ref{tbl-PM}}
\tablenotetext{b}{Final $\pi_{abs}$ from modeling {\it HST} data with equations 2--5, employing the constraints 
summarized in Section~\ref{Priors}}
\tablenotetext{b}{$V_t = 4.74\times \mu/\pi_{abs}$}
\end{deluxetable}

\begin{deluxetable}{rl}\tablewidth{0in}
\tablecaption{Pleiades Parallax and Proper Motion \label{tbl-SUM}}
\tablewidth{0in}
\tablehead{\colhead{Parameter} &  \colhead{ Value }}
\startdata
{\it HST} study duration  &3.51 y \\
number of observation sets    &   6  \\
reference star $\langle V\rangle$ &  $14.63 $   \\
reference star $\langle (B-V) \rangle$ &  $0.9\tablenotemark{a} $  \\
 \\
{\it HST} Absolute Parallax \tablenotemark{b}  & \HPI     mas \\
{\it HST} Relative Proper Motion \tablenotemark{c}  & $46.3\pm3.7$  mas y$^{-1}$  \\
 \indent in pos. angle & 158\arcdeg  $\pm1$\arcdeg   \\
\enddata
\tablenotetext{a}{Estimated from {\it VJHK} photometry and spectral types, with $A_V = 0.14$ for 
dwarfs and $A_V = 0.23$ for giants. }
\tablenotetext{b}{Average of 3030, 3063, and 3179 from Table~\ref{tbl-PiVt}}
\tablenotetext{c}{Average of 3030, 3063, and 3179 from Table~\ref{tbl-PM}, `Final.' Proper motion 
error is the standard deviation of the individual measures.}
\end{deluxetable}

\begin{deluxetable*}{lccccl}\tablewidth{0in}
\tablecaption{Previous and Present Pleiades Parallaxes \label{tbl-Prev}}
\tablewidth{0in}
\tablehead{\colhead{Method} & \colhead{Abbr.} &  \colhead{ $\pi_{abs}$ } & \colhead{$d$ (pc)} & \colhead{$(m-M)$} &  \colhead{Reference}}
\startdata
{\it HST}/FGS parallax & {\it HST} & $7.43\pm0.17$  & $134.6\pm3.1$ & $5.65\pm0.05$ &this paper \\
Hipparcos all-sky & HIP & $8.45\pm0.25$ & $118.3\pm3.5$ & $5.37\pm0.06$ &\citet{vanL99} \\
Allegheny Obs. & AO & $7.64\pm0.43$ & $130.9\pm7.4$ & $5.59\pm0.11$ &\citet{Gat00} \\
interferometric orbit & Pan & $7.41\pm0.11$ & $135.0\pm2.0$ & $5.65\pm0.03$ &\citet{Pan04} \\
dynamical parallax & Mun & $7.58\pm0.11$ & $131.9\pm3.0$ & $5.60\pm0.05$ &\citet{Mun04} \\
main sequence fitting & MS & $7.58\pm0.14$ & $131.9\pm2.4$ & $5.60\pm0.04$ &\citet{Pin98} \\
\enddata
\end{deluxetable*}

\begin{deluxetable*}{lrrl}
\tablewidth{0in}
\tablecaption{{\it HST} and {\it Hipparcos} Absolute Parallaxes    \label{tbl-HH}}
\tablehead{\colhead{Object}& \colhead{$\pi_{HST}$ (mas)}& \colhead{$\pi_{\it HIP}$ (mas)} &\colhead{{\it HST} Reference}}
\startdata
Proxima Cen     & $769.7\pm0.3\phn$  & $772.33\pm2.42$ & \citet{Ben99} \\
Barnard's Star   & $545.5\pm0.3\phn$  & $549.3\pm1.58$  &  \citet{Ben99} \\
Gliese 876         & $214.6\pm0.2\phn$  & $212.7\pm2.1\phn$    &  \citet{Ben02c} \\
Feige 24             & $14.6\pm0.4\phn$    & $13.44\pm3.62$  &  \citet{Ben00} \\
Wolf 1062           & $98.0\pm0.4\phn$    & $98.56\pm2.66$  &  \citet{Ben01} \\
Pleiades             & $7.43\pm0.17$  &   $8.45\pm0.25$  & this paper \\
RR Lyrae            & $3.60\pm0.20$  &   $4.38\pm0.59$  &  \citet{Ben02a} \\
$\delta$ Cephei & $3.66\pm0.15$ &   $3.32\pm0.58$  &  \citet{Ben02b} \\
HD 213307         & $3.65\pm0.15$ &   $3.43\pm0.64$  &  \citet{Ben02b} \\
\enddata
\end{deluxetable*}

\begin{figure}
\epsscale{0.85}
\plotone{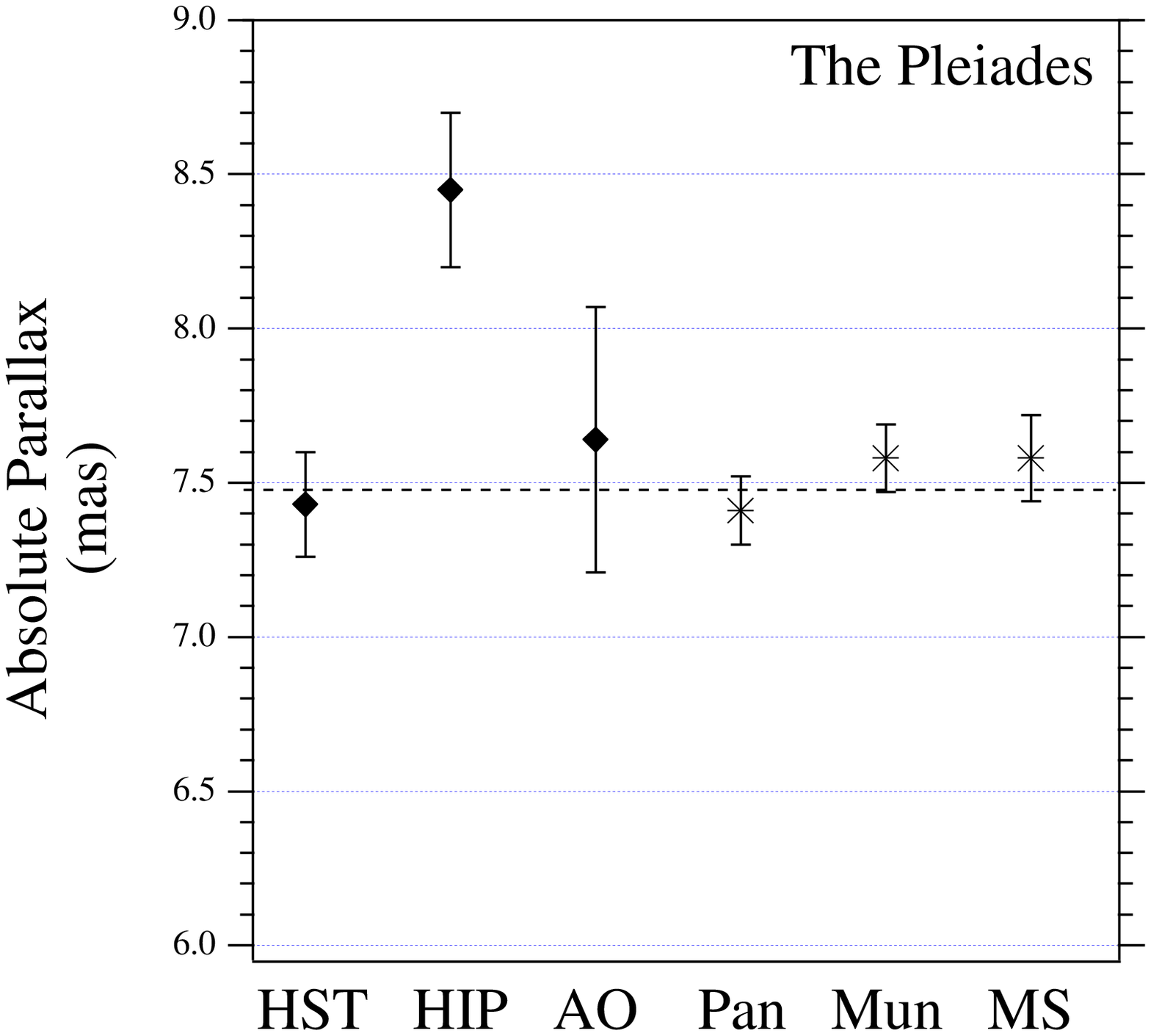}
\caption{Absolute parallax determinations for the Pleiades. We compare astrometric parallax results (filled diamonds) from {\it HST} (``{\it HST}"), {\it Hipparcos} (``Hip"), andrecent determinations from Allegheny Observatory (``AO"), \citet{Gat00}.  \citet{Pan04} (``Pan") have derived a dynamical parallax from long-baseline interferometryand radial velocity measurements of the binary star Atlas. \citet{Mun04} have performeda similar dynamical determination on another Pleiades binary.  MS denotes a parallax derived from main-sequence fitting \citep{Pin98}. The horizontal dashed line is the weighted average of the {\it HST}, Pan, Munari and AO measures.}
\label{fig-5}
\end{figure}

Note that we do not measure the parallax of these Pleiads relative to a reference frame with unknown parallax and then apply a correction to absolute parallax, assuming some model of the Galaxy.  In a quasi-Bayesian approach, the reference star spectrophotometric absolute parallaxes, UCAC2 and \citet{Sch95} proper motions, and an estimated cluster depth were input as observations with associated errors, not as hard-wired quantities known to infinite precision.  Parallaxes and relative proper motion results from {\it HST} are collected in Tables~\ref{tbl-PM} and \ref{tbl-PiVt}. We obtain for the Pleiades members an average absolute parallax $\pi_{abs} =$ \HPI mas (Table~\ref{tbl-SUM}).  Because we employ a cluster depth constraint, the three Pleiades member parallaxes are not independent measurements. Hence, we cannot use the standard deviation of the mean to reduce our final error by $\sqrt{2}$.  Along with our result, other recent Pleiades parallaxes are listed in Table~\ref{tbl-Prev} and compared in Figure~\ref{fig-5}. The most discrepant of these is clearly and only the {\it Hipparcos} result. 

Our absolute parallax for the Pleiades contains one last systematic uncertainty: where in the cluster do our three Pleiades members lie? In Section~\ref{Priors} we estimated a `depth' in parallax of $\sim0.20$ mas. Our final parallax result should be stated $\pi_{abs} =$ \HPI $\pm$ 0.20 mas with the error having both a random and systematic component. We point out that each of the astrometric results in Figure~\ref{fig-5} suffers from the same systematic error. In the next Section we reduce that error by averaging those results.  Inspecting Tables~\ref{tbl-PM} and \ref{tbl-PiVt} we note that ref-14, identified as Cl* Melotte 22 CALAR 7, is in fact not a Pleiad, disagreeing in parallax and proper motion with the first three stars in these Tables, all identified members.

\section{Discussion and Summary}

\subsection{{\it HST} Parallax Accuracy}
\begin{figure}
\plotone{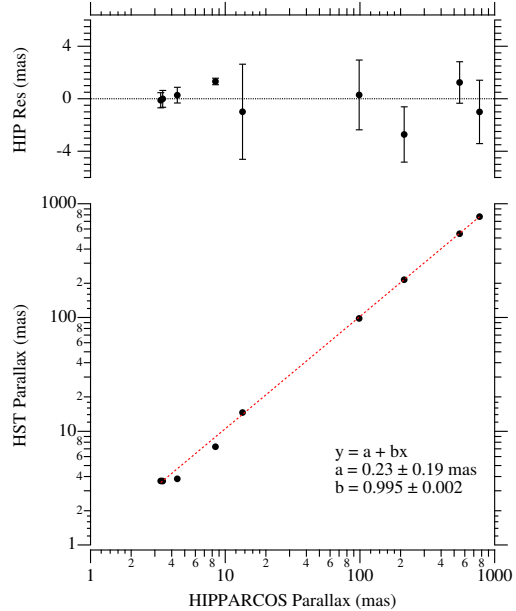}
\caption{ Bottom: {\it HST} absolute parallax determinations compared
with {\it Hipparcos} for all targets listed in Table \ref{tbl-HH}. Top: The {\it Hipparcos} residuals to the dotted error-weighted impartial regression line that excludes the Pleiades. The error bars on the residuals are {\it Hipparcos} Catalog $1\sigma$ errors. }
\label{fig-6}
\end{figure}

Our parallax precision, an indication of our internal, random error, is often less than 0.3 mas. To assess our accuracy, or external error, we must compare our parallaxes with results from independent measurements. Following \citet{Gat98} and extending the analysis presented in \citet{Ben02b} with the addition of a recent parallax for Gl 876 \citep{Ben02c}, we plot eight parallaxes obtained by the {\it HST} Astrometry Science Team with FGS 3 and, now FGS 1r, against those obtained by {\it Hipparcos}. Data for these objects are collected in Table \ref{tbl-HH} and shown in Figure \ref{fig-6}. The dashed line is a weighted regression that takes into account errors in both input data sets and excludes the Pleiades. Figure~\ref{fig-6} indicates no statistically significant scale difference compared to {\it Hipparcos}. However, for this fit, which excludes the Pleiades, we obtain a reduced $\chi^2 = 0.265$. Including the Pleiades, we obtain a significantly poorer fit with reduced $\chi^2 = 0.551$, again, suggesting a problem with the {\it Hipparcos} Pleiades parallax.

Our result, in and of itself, does not lead to the conclusion that the {\it Hipparcos} parallax for the Pleiades is wrong, but that conclusion cannot be avoided once all the results are examined together.  Especially important for making this case are the two recent determinations of visual binary orbits for Pleiades members.  \citet{Pan04} used the Palomar Testbed Interferometer to determine very precise relative positions of the two stars comprising Atlas, one of the Seven Sisters.  Without having a radial velocity orbit they could not determine all the parameters, but a solution is possible by assuming masses for the stars, and the masses enter in the cube root.  By doing this they concluded that the distance to the Pleiades cannot be less than 127 pc and that the most likely distance lies between 133 and 137 pc.  \citet{Mun04} analyzed light- and radial velocity curves for HD 23642, an eclipsing binary in the Pleiades, and determined a distance of $132\pm2$ pc.  (This would decrease to $130.6\pm3.7$ if the assumed reddening were increased to as much as $E(B-V) = 0.035$ magnitude).

\subsection{The Distance to the Pleiades}

There now exist three completely independent determinations of the Pleiades distance that use completely independent techniques and data, and they all yield the same answer to within their errors.  Our traditional parallax determination leads to $d = 134.6\pm3.1$ pc, a visual binary orbit leads to $d = 135\pm2$, and an eclipsing binary orbit results in $d = 132\pm2$.  For comparison, recent estimates from main sequence fitting include $132\pm4$ \citep{Ste01} and $132\pm2$ \citep{Pin98}, and \citet{Gat00} has determined $131\pm7$ at Allegheny Observatory.  \citet{Nar99} derived $131\pm24$ pc from the gradient in the radial velocities of Pleiades members in the direction of the cluster's proper motion.  

Clearly the {\it Hipparcos} result, $118\pm4$, is discrepant.  This can be seen graphically in Figure~\ref{fig-6}, and a summary of these distance determinations is given in Table~\ref{tbl-Prev}.  Understanding this discrepancy is crucial.  Astrometry is arguably the one branch of astronomy where accurate and precise knowledge of uncertainties cannot be overlooked.  The {\it Hipparcos} team has been well aware of this and has put considerable effort into examining potential sources of systematic error.  Their most recent papers (\citet{vLF03a}, \citet{DTvL03}, \citet{vLP03}, \citet{FvL03}, \citet{vLF03b}) show, for instance, that noise in the along-scan attitude dominates for $H_p < 4.5$ (where $H_p$ is the apparent magnitude as directly measured by {\it Hipparcos}) and that this may be especially important for the Pleiades, inter alia \citep{vLF03b}.  This possibility was examined by \citet{Mak02}, who reanalyzed {\it Hipparcos} data to derive $d = 129\pm3$, a value that is substantially less discrepant than that reported by \citet{vanL99} and \citet{Rob99}.

The answer certainly does not lie in an unusual shape or physical properties for the Pleiades.  \citet{Ste01} suggested that the {\it Hipparcos} distance could be reconciled with traditional measures if the bright stars---the Seven Sisters---that dominate the {\it Hipparcos} result happen to lie at the near end of an elongated cluster.  This is disproved by the fact that \citet{Pan04} find Atlas itself to lie at the traditional distance.  \citet{Gre01} suggested that the luminosities of Pleiades stars could be accounted for by a low cluster metallicity of $-0.112\pm0.025$, determined from Geneva photometry.  The exact metallicity of the Pleiades remains uncertain, but it is unlikely to be as low as that since analyses from high-resolution spectra yield values that are essentially solar (e.g., \citet{Boe90} get [Fe/H] = $-0.034\pm0.024$).  \citet{Hai01} have likewise refuted the \citet{Gre01} metallicity on several grounds.

\bigskip

To summarize, {\it HST} astrometry yields an absolute trigonometric parallax for three members of the Pleiades, $\pi_{abs} = $\HPI mas with a 0.20 mas systematic error due to cluster depth. A weighted average with previous ground-based astrometric determinations ({\it HST}, AO, Pan and Munari, Table~\ref{tbl-Prev}) provides $\pi_{abs} = 7.49\pm0.07$ mas. This average result should reduce the contribution of the cluster depth systematic error, presuming that the stars measured by these techniques are randomly distributed within the cluster. With ${\sigma_{\pi} \over \pi}\sim1$\%, any Lutz-Kelker-Hanson bias correction (\citet{Lut73}, \citet{Han79}) to an absolute magnitude would be less than 0.01 magnitude (e.g., \citet{Ben02b}).  This net parallax of $7.49\pm0.07$ mas corresponds to $d = 133.5\pm1.2$ pc, or $(m-M) = 5.63\pm0.02$ magnitude.  This is likely to be the best available distance for the Pleiades until observations of substantially better precision can be made with a mission such as SIM or GAIA.

\bigskip

\acknowledgments

We thank an anonymous referee for comments which materially improved the clarity of our presentation. Support for this work was provided by NASA through grant GO-08335 from the Space Telescope Science Institute, which is operated by AURA, Inc., under NASA contract NAS5-26555.  These results are based partially on observations obtained with the Lick Observatory 3 m telescope, which is owned and operated by the University of California.  This publication makes use of data products from the Two Micron All Sky Survey, which is a joint project of the University of Massachusetts and the Infrared Processing and Analysis Center/California Institute of Technology, funded by NASA and the NSF.  This research has made use of the SIMBAD database, operated at CDS, Strasbourg, France; the NASA/IPAC Extragalactic Database (NED) which is operated by JPL, California Institute of Technology, under contract with the NASA;  and NASA's Astrophysics Data System Abstract Service.



%

\end{document}